\documentclass[conference]{IEEEtran}

\IEEEoverridecommandlockouts

\usepackage{cite}
\usepackage{amsmath,amssymb,amsfonts}
\usepackage{algorithmic}
\usepackage{graphicx}
\usepackage{textcomp}

\usepackage{booktabs, multirow}
\usepackage{subcaption}
\usepackage[table]{xcolor}
\usepackage{amssymb}
\usepackage{url}
\usepackage{booktabs, multirow}
\usepackage{multirow}
\usepackage{graphicx}
\usepackage[utf8]{inputenc}
\usepackage[T1]{fontenc}
\usepackage{xcolor}
\usepackage{comment}
\usepackage{amssymb}
\usepackage{amsmath}
\usepackage{booktabs, multirow}
\usepackage{subcaption}
\usepackage{pifont}
\usepackage{array, makecell}
\usepackage{url}
\usepackage{mathrsfs}

\usepackage[colorlinks=true,linkcolor=cyan,urlcolor=red]{hyperref}
\hypersetup{
    colorlinks=true,
    linkcolor=black,
    filecolor=magenta,
    citecolor=black,
    urlcolor=black,
    }

\usepackage[numbers]{natbib}
\def\BibTeX{{\rm B\kern-.05em{\sc i\kern-.025em b}\kern-.08em
    T\kern-.1667em\lower.7ex\hbox{E}\kern-.125emX}}
\begin{document}

\title{Noise-Agnostic Multitask Whisper Training for Reducing False Alarm Errors in Call-for-Help Detection}

\author{\IEEEauthorblockN{Myeonghoon Ryu$^{*\dagger1}$\thanks{\hspace{-1em}$^{*}$equal contributions $^{\dagger}$ corresponding authors, This study was supported by the Korea Health Technology R\&D Project through the Korea Health Industry Development Institute (KHIDI), funded by the Ministry of Health \& Welfare, Republic of Korea (Grant HI22C1962), and the Mental Health related Social Problem Solving Project, Ministry of Health \& Welfare, Republic of Korea (Grant RS-2024-00403474). This research was supported by Brian Impact Foundation, a non-profit organization dedicated to the advancement of science and technology for all.}, June-Woo Kim$^{*\dagger2,3}$, Minseok Oh$^{1}$, Suji Lee$^{1}$, Han Park$^{1}$}
\IEEEauthorblockA{\textit{$^{1}$Deeply Inc., $^{2}$Department of Artificial Intelligence, Kyungpook National University, $^{3}$RSC LAB, MODULABS} \\
{$^{1}$Seoul, South Korea; $^{2}$Daegu, South Korea, $^{3}$Seoul, South Korea} \\
myeonghoon.ryu@deeply.co.kr; kaen2891@gmail.com}
}

\maketitle

\begin{abstract}
Keyword spotting is often implemented by keyword classifier to the encoder in acoustic models, enabling the classification of predefined or open vocabulary keywords. Although keyword spotting is a crucial task in various applications and can be extended to call-for-help detection in emergencies, however, the previous method often suffers from scalability limitations due to retraining required to introduce new keywords or adapt to changing contexts. We explore a simple yet effective approach that leverages off-the-shelf pretrained ASR models to address these challenges, especially in call-for-help detection scenarios. Furthermore, we observed a substantial increase in false alarms when deploying call-for-help detection system in real-world scenarios due to noise introduced by microphones or different environments. To address this, we propose a novel noise-agnostic multitask learning approach that integrates a noise classification head into the ASR encoder. Our method enhances the model's robustness to noisy environments, leading to a significant reduction in false alarms and improved overall call-for-help performance. Despite the added complexity of multitask learning, our approach is computationally efficient and provides a promising solution for call-for-help detection in real-world scenarios. 

\end{abstract}


\section{Introduction}
\label{sec:intro}
ASR technology has emerged as a pivotal component in contemporary human-computer interaction, enabling applications ranging from virtual assistants to assistive solutions for individuals with hearing impairments. Among the various sub-tasks in ASR, keyword spotting (KWS) has gained considerable attention due to its critical role in enabling voice-activated systems~\cite{chen2014small, hoy2018alexa}. While previous KWS studies has shown promise in various applications~\cite{chen2014small, sacchi2019open, liu2021rnn, zhang2023u2, nishu2023matching}, its effectiveness in detecting call-for-help in hazardous environments remains underexplored. We aim to address this gap by leveraging ASR and KWS concepts to develop an effective system for call-for-help detection in real-world scenarios~\cite{chu2021call}.

KWS specifically involves detecting predefined words or phrases within a continuous speech stream, serving as a trigger for executing predetermined actions, such as activating smart speakers or initiating command sequences~\cite{warden2018speech}. Typically, a keyword classification head is attached to the encoder of an acoustic model to identify predefined keywords. While this method has demonstrated efficacy in finite keyword settings, such as query-by-example~\cite{chen2015query, settle2017query, kim2019query, huang2021query, yang2021superb}, it frequently encounters challenges when deployed in real-world environments. One key limitation is scalability. Previous approaches require retraining when new keywords are introduced or when the system needs to adapt to different contexts. This becomes particularly problematic in dynamic environments where new keywords may emerge. Consequently, this constant retraining can lead to inefficiencies for KWS service.
To address these issues, studies on \textit{Open-Vocabulary} KWS system were emerged~\cite{mo2021encoder, liu2021rnn, berg2021keyword, zhang2023u2}, including few-shot trials~\cite{mazumder2021few}, leveraging speech embeddings~\cite{kirandevraj2022generalized} and text-audio embeddings~\cite{sacchi2019open, shin2022learning, nishu2023matching}. Despite advancements in scalability, their works often involve complex network architectures and rely on cascaded ASR systems for KWS, which can introduce challenges for user implementation.

This paper proposes a simple yet effective approach for KWS, particularly focused on detecting \textit{calls-for-help} in emergency situations. By fine-tuning an off-the-shelf large-scale Whisper~\cite{radford2023robust}, on a small dataset specific to emergency-related calls for help, we aim to develop a more adaptive and scalable KWS system. Our method leverages ASR outputs to enable quick adaptation to new keywords without the need for retraining, making it highly suitable for real-world deployment.

However, deploying systems in real-world environments is a challenging tasks, especially due to noise~\cite{hubballi2014false, chen2015query, zhang2017hello, huang2021query, ng2022i2cr}. Microphones in devices like smartphones and smart speakers often pick up substantial background noise, which can significantly impair system performance. For the call-for-help in emergency situations, where both accuracy and speed of ASR are critical, false alarms triggered by noisy conditions can have serious consequences.

We introduce a multitask learning approach that integrates a noise classification head into the Whisper encoder. This multitask setup enables the model to learn noise patterns that commonly trigger false alarms, enhancing its robustness in noisy environments while using only a small number of additional model parameters. By simultaneously training the model to recognize both keywords and environmental noise types, our model can be a noise-agnostic call-for-help detection system in practical scenarios. Experimental results validate that our proposed method improves the model's robustness in noisy environments, significantly reducing false alarms and enhancing overall performance in real-world scenarios. Our contributions are summarized below:

\begin{itemize}
\item We present a simple yet effective approach to call-for-help detection in emergency situations. By leveraging off-the-shelf Whisper for fine-tuning, we address the scalability issues present in previous KWS systems.

\item We propose a multitask learning approach for a noise-agnostic call-for-help detection system that requires only a few additional model parameters. Our method significantly reduces false alarms while enhancing overall performance in noisy environments, offering a practical solution for emergency response.

\item To validate the effectiveness of our method in real-world scenarios, we recruited participants and recorded call-for-help situations. As a result, our method significantly reduced false alarms in diverse environments. Finally, we provide the code to the KWS community to support further research and development.

\end{itemize}
\section{Method}

\subsection{Fine-tuning Whisper for call-for-help detection}
Whisper~\cite{radford2023robust} is a large-scale pretrained ASR model trained on 680K hours of multilingual speech data. Besides, Whisper is further trained on various speech processing tasks using a multitask supervision approach, such as translation, voice activity detection, and language identification, etc. By leveraging the pretrained Whisper model, it can be easily fine-tuned to extend its rich knowledge for several downstream tasks.

\begin{figure}[!t]
    \centering
    \includegraphics[width=1.0\linewidth]{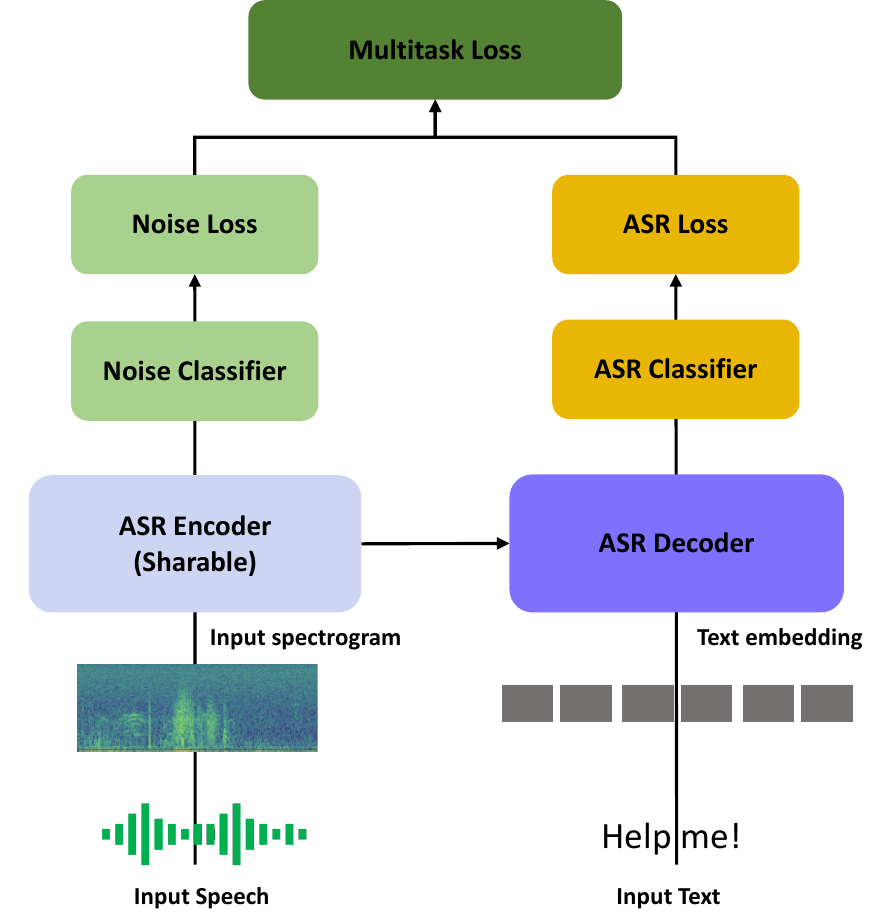}
    \caption{Overall illustration of proposed multitask learning approach that integrates a noise classification head into the Whisper encoder.}
    \label{fig:fig1}
    \vspace{-5mm}
\end{figure}

We use the Whisper model for fine-tuning on call-for-help detection tasks, which are closely related to keyword spotting. Unlike previous approaches that rely heavily on predefined keywords or acoustic models, we categorize call-for-help scenarios and utilize ASR results for classification. In this work, we define emergency scenarios as: 1) \textbf{shouting for help in life-threatening situations} 2) \textbf{crying out for assistance when personal safety is at risk} and 3) \textbf{other similar situations}, though not limited to these. Based on the speech and its corresponding text, e.g., \textbf{help me} or \textbf{save me}, we classify these as emergency tasks. All other results are categorized as \textbf{others}.

\subsection{Multitask training for reducing false alarm errors}
To alleviate the false alarm errors in noisy real-world scenarios, we introduce multitask learning for noise classification. As shown in Fig.~\ref{fig:fig1}, noise classifier is added to the Whisper encoder, therefore the ASR encoder is sharable for noise classification as well as speech recognition. Therefore, the final loss $\mathcal{L}_{\text{Multi}}$ can be formularized as:

\begin{figure*}[!t]
    \centering
    \includegraphics[width=1.0\linewidth]{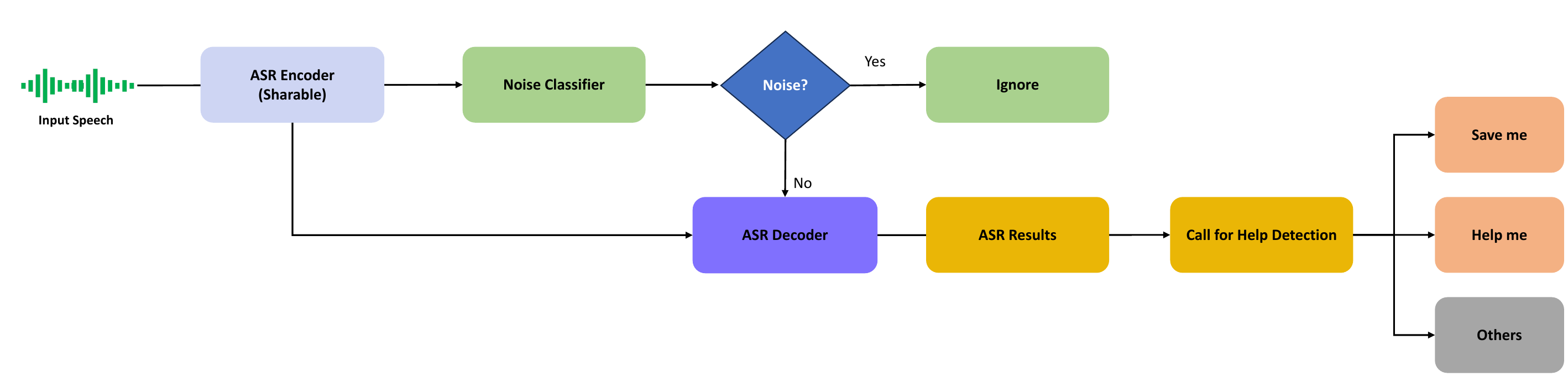}
    \caption{Overview of the proposed call-for-help detection system using our approach in real-world scenarios.}
    \label{fig:fig2}
    \vspace{-3mm}
\end{figure*}

\vspace{-1mm}
\begin{align}\label{eq:sdat1} 
\mathcal{L}_\text{Multi} = \mathcal{L}_\text{Noise} + \mathcal{L}_\text{Seq2Seq}
\end{align}
where the noise loss $\mathcal{L}_\text{Noise}$ and ASR loss $\mathcal{L}_\text{Seq2Seq}$ are:
\vspace{-1mm}
\begin{align}\label{eq:sdat1} 
\mathcal{L}_{\text{Noise}} \! = -\sum_{i=1}^n\! \, m_{i}\! \, \log \, \!(\hat{m_{i}}),
\end{align}
\vspace{-1mm}
\begin{align}\label{eq:sdat1} 
\mathcal{L}_{\text{Seq2Seq}} \! = -\sum_{t=1}^T\! \, \log P(y_t|y_1,...,y_{t-1},\mathbf{x}).
\end{align}
Here, both $\mathcal{L}_\text{Noise}$ and $\mathcal{L}_\text{Seq2Seq}$ are based on Cross-Entropy loss, with multitask noise label $m$ and ASR label $y$ (division by $N$ and $T$ are omitted). The predicted probabilities $\hat{m}$ and $\hat{y}$ are generated by passing through the noise classifiers and ASR decoder, followed by the ASR classifier (i.e., with the same vocabulary length). While general multitask learning techniques often involve complex architectures or bottleneck issues, our approach remains simple, requiring only a minimal increase in model parameters. For instance, when using the Whisper-tiny model, just 384 hidden units are added.

\subsection{Evaluation stage for call-for-help detection}
To mitigate the computational costs in real-world scenarios, our evaluation stage for call-for-help detection is illustrated in Fig.~\ref{fig:fig2}. The proposed system commences by acquiring audio data from a microphone. First, input waveform from microphones are resampled and converted to the Mel-spectrogram format. The output audio representations from ASR encoder are then fed into a noise classifier to determine the presence of noise within the speech signal. In the event of noise detection, the system terminates processing, thereby conserving computational resources when the noise predictor exhibits high accuracy. Conversely, if speech is identified, the extracted audio features are forwarded to the ASR decoder, adhering to conventional ASR protocols. Subsequently, the transcribed output is categorized according to predefined emergency tasks, facilitating the detection of call-for-help requests.
\section{Experiments}

\begin{figure*}[!t]
    \vspace{-1mm}
    \centering
    \begin{subfigure}{.5\linewidth}
      \centering
      \includegraphics[width=1.0\linewidth]{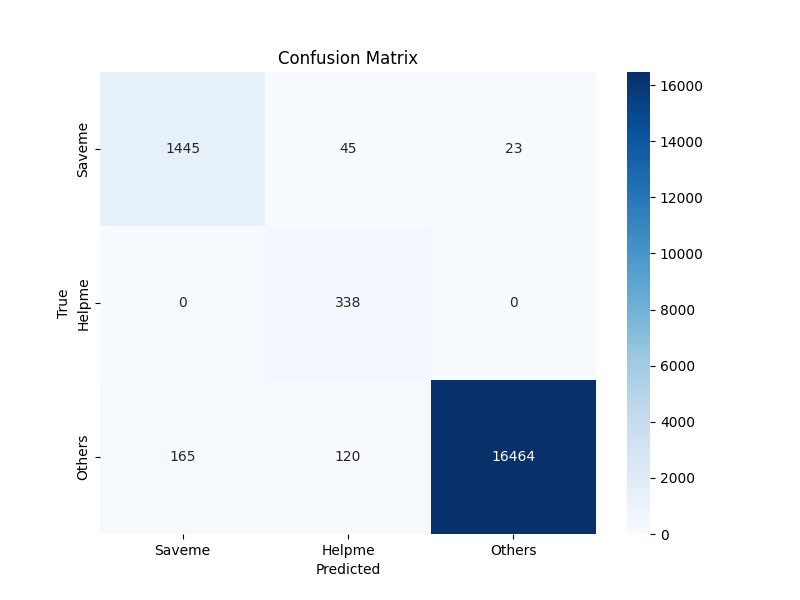}
      \caption{Whisper fine-tuned (3 class).}
      \label{fig:sfig1}
    \end{subfigure}%
    \begin{subfigure}{.5\linewidth}
      \centering
      \includegraphics[width=1.0\linewidth]{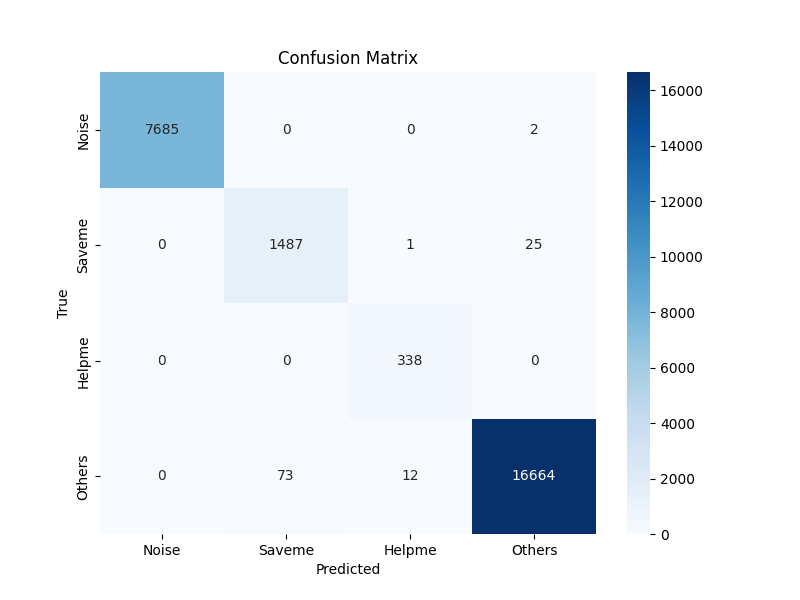}
      \caption{Whisper multitask (4 class).}
      \label{fig:sfig2}
    \end{subfigure}    
    
    \caption{Confusion matrices of both whisper fine-tuned and multitask approach.}
    \vspace{-5mm}
    \label{fig:confusion}
    \end{figure*}

\subsection{Setting}
\textbf{Emergency dataset:} We used the public Korean Emergency Audio and Speech dataset\footnote{\url{https://www.aihub.or.kr/aihubdata/data/view.do?currMenu=115&topMenu=100&aihubDataSe=data&dataSetSn=170}}, which consists of 16 emergency situations and approximately 3,500 hours of 44 kHz audio. The dataset includes 422K training and 55K test samples across various emergency scenarios, such as call-for-help, fire, gas accidents, forced harassment, theft, robbery, and violent crime. In this study, we focused on six call-for-help scenarios, leading to final training and test sets comprising 186K and 18K samples. Specifically, for the keywords \textbf{saveme} and \textbf{helpme}, the training set contains 5,250 and 2,720 samples, while the test set includes 338 and 1,513 samples, respectively.

\textbf{Noise dataset:} For multitask learning with noise classification, we employed CochlScene~\cite{jeong2022cochlscene} acoustic scene dataset, which consists of 76K samples collected across 13 distinct acoustic scenes. We also consider the MS-SNSD~\cite{reddy2019scalable} dataset for out-of-domain noise classification performance evaluation.

\textbf{Real-world recording dataset:} We aimed to collect audio data that accurately represents typical real-world environments, focusing on tasks such as ASR and call-for-help detection. We collected a total of 547 audio samples, comprising both speech data and background noise. The speech data comprised 202 recordings of the phrase \textbf{saveme} (six distinct utterances in Korean) and 145 recordings of the phrase \textbf{helpme} (three distinct utterances in Korean), spoken by eight participants with an equal gender ratio in diverse environments, including indoor spaces, outdoor areas, and public restrooms. We assume that the recorded samples have an SNR of approximately 20 dB. To ensure a comprehensive evaluation of call-for-help detection model in noisy conditions, we also gathered 200 background noise recordings from offices, restrooms, and public spaces. All samples were captured using \textbf{Spon TS-905E} (omnidirectional), \textbf{Spon TS-715E} (directional and omnidirectional), and \textbf{Spon TS-929E} (unidirectional) microphones.

\textbf{Training details:} We fine-tuned the Whisper-tiny model for 10 epochs, employing the AdamW optimizer with a learning rate of 1e--4, cosine learning rate scheduling, and a batch size of 32. For multitask noise classification training, the output of the Whisper encoder was processed by averaging along the time axis. To ensure stable training, momentum updates with a coefficient of 0.5 were applied to all learnable parameters. This fine-tuning setting was conducted under both single-task and multitask learning paradigms.

\textbf{Metrics:} We report the call-for-help detection accuracy across three classes (\textit{saveme}, \textit{helpme}, \textit{others}). We also provide results for noise classification accuracy and F1-score.

\subsection{Results}

\begin{table}[!t]
    \centering
    \caption{
    Comparison of call-for-help accuracy under different pretrained models.
    \textbf{Bold} denotes the best result.}\label{tab:table1}
    \renewcommand{\arraystretch}{1}
    \addtolength{\tabcolsep}{1pt}
    \resizebox{\linewidth}{!}{
    \begin{tabular}{lllll}

    \toprule
    Model & \# Params & Pretrained With & Accuracy\,(\%) \\
    \hline \midrule

    EfficientNet\, \cite{tan2019efficientnet} & 5.3M & Image & 80.44 \\
    CNN6\, \cite{kong2020panns} & 4.8M & Image & 81.52 \\
    AST\, \cite{gong2021ast} & 87.7M & Image + Audio & 82.28 \\

    \hline \midrule

    wav2vec2-Base\, \cite{baevski2020wav2vec} & 95M & Read Speech & 94.30 \\
    HuBERT-Large\, \cite{hsu2021hubert} & 310M & Read Speech & 95.32 \\
    XLSR300M\, \cite{babu2021xls} & 310M & Read Speech & 95.20 \\
    Whisper-Tiny\, \cite{radford2023robust} & 39M & Multiple Speech  & \textbf{98.54} \\

    \bottomrule

    \end{tabular}
    }

\end{table}
\textbf{Effectiveness of Whisper:} To evaluate the Whisper model's performance relative to other networks for call-for-help detection, we compare with EfficientNet~\cite{tan2019efficientnet}, CNN6~\cite{kong2020panns}, Audio Spectrogram Transformer~\cite{gong2021ast}, wav2vec2~\cite{baevski2020wav2vec} Base, Hubert~\cite{hsu2021hubert} Large, and XLSR300M~\cite{babu2021xls}. Note that these models were trained under predefined KWS settings, specifically tailored for classification tasks as described in~\cite{yang2021superb}. Our choice of various pretrained models allows us to better understand the impact of pretraining data quantity, out-of-domain embeddings, and model parameter counts on performance.

As shown in Table~\ref{tab:table1}, out-of-domain models~\cite{tan2019efficientnet, kong2020panns, gong2021ast} struggle with call-for-help detection. In contrast, self-supervised speech models achieved better results. Our findings indicate that the number of parameters does not highly related to call-for-help detection performance. Although self-supervised speech models are trained on classification layer--simpler than ASR fine-tuning--the Whisper model outperformed these pretrained speech models despite having fewer parameters. This suggests that the Whisper model is particularly well-regarded for call-for-help detection and could be extended to additional keywords based on its ASR performance.

\textbf{Effectiveness of multitask learning:} Fig.~\ref{fig:confusion} illustrates the confusion matrices for Whisper fine-tuning and multitask learning results. 
Interestingly, the overall performance of the multitask method surpassed that of Whisper fine-tuning alone, suggesting that incorporating a simple noise classifier into the Whisper encoder for multitask learning allows the model to benefit from the additional task, specifically improving noisy robustness, and leading to enhanced performance.

\textbf{Overall performance on recording samples:} Table~\ref{tab:table2} presents the call-for-help detection results in real-world scenarios, specifically using recordings from our microphones. The detection accuracy and F1-score are reported for the vanilla Whisper-tiny model, the Whisper fine-tuned model, and the multitask learning approach. As expected, the proposed multitask learning method significantly outperformed the approach without noise classification training, indicating that incorporating noise training contributes to more reliable and stable call-for-help detection based on ASR outputs.


\begin{table}[!t]
    \centering
    \caption{Call-for-help performance on recording samples according to different methods.}\label{tab:table2}
    \vspace{-1mm}
    \renewcommand{\arraystretch}{1}
    \addtolength{\tabcolsep}{5pt}
    \resizebox{\linewidth}{!}{
    \begin{tabular}{l|ll}

    \toprule
    Method & Accuracy\,(\%) & F1-score (Macro) \\
    \hline \midrule

    Vanila Whisper & 56.86\ & 0.53 \\
    Whisper fine-tuned & 65.03 & 0.63 \\
    Whisper multitask & 88.48\ & 0.89 \\
    \bottomrule

    \end{tabular}
    
    \vspace{-5mm}
    }
\end{table} 

\begin{table}[!h]
    \centering
    \caption{Noise classification performance on in-domain and out-of-domain dataset.}\label{tab:table3}
    \renewcommand{\arraystretch}{1}
    \addtolength{\tabcolsep}{20pt}
    \resizebox{\linewidth}{!}{
    \begin{tabular}{ll}

    \toprule
    Data type & Accuracy\,(\%) \\
    \hline \midrule

    In-domain test set \cite{jeong2022cochlscene} & 98.43\ \\
    Out-of-domain test set \cite{reddy2019scalable} & 58.82\ \\
    \bottomrule

    \end{tabular}
    \vspace{-5mm}
    }
\end{table}

\textbf{Discussions:}
To evaluate the generalizability of the proposed multitask learning approach for noise classification, we report accuracy on both in-domain and out-of-domain datasets. The in-domain dataset refers to the CochlScene~\cite{jeong2022cochlscene} test set (7,687 samples), while the out-of-domain dataset corresponds to the MS-SNSD~\cite{reddy2019scalable} test set (51 samples), which includes various environments, machine noises, and background speech sounds. As shown in Table~\ref{tab:table3}, our method achieved a noise classification accuracy of 98.43\% across the test sets. However, performance on the out-of-domain dataset (i.e., unseen data) was significantly lower compared to the in-domain test set, indicating that the model still struggles with noise classification in diverse noisy environments and when recordings are captured using different microphones. We hypothesize that the lower accuracy is affected by the presence of background human speech in the MS-SNSD dataset. To address this, few-shot fine-tuning using recordings from varied environments or microphones, as well as domain adaptation techniques~\cite{ganin2016domain} for device-specific issues~\cite{kim2024stethoscope} or augmentation~\cite{park2019specaugment, kim2024repaugment}, should be considered.
\section{Conclusion}
This paper focused on a novel KWS-based system designed to accurately identify call-for-help detection in real-world scenarios characterized by challenging noisy acoustic conditions. To this end, we proposed a multitask learning framework for a noise-robust system. Our empirical findings demonstrated that this approach is both straightforward and effective for call-for-help detection in real-world environments. Nevertheless, future research will be required to enhance the system's noise classification capabilities across a broader spectrum of scenarios.


\small
\bibliography{refs}

\begin{thebibliography}{10}

\bibitem{babu2021xls}
A.~Babu, C.~Wang, A.~Tjandra, K.~Lakhotia, Q.~Xu, N.~Goyal, K.~Singh, P.~von Platen, Y.~Saraf, J.~Pino, et~al.
\newblock Xls-r: Self-supervised cross-lingual speech representation learning at scale.
\newblock {\em arXiv preprint arXiv:2111.09296}, 2021.

\bibitem{baevski2020wav2vec}
A.~Baevski, Y.~Zhou, A.~Mohamed, and M.~Auli.
\newblock wav2vec 2.0: A framework for self-supervised learning of speech representations.
\newblock {\em Advances in neural information processing systems}, 33:12449--12460, 2020.

\bibitem{berg2021keyword}
A.~Berg, M.~O'Connor, and M.~T. Cruz.
\newblock Keyword transformer: A self-attention model for keyword spotting.
\newblock {\em arXiv preprint arXiv:2104.00769}, 2021.

\bibitem{chen2014small}
G.~Chen, C.~Parada, and G.~Heigold.
\newblock Small-footprint keyword spotting using deep neural networks.
\newblock In {\em 2014 IEEE international conference on acoustics, speech and signal processing (ICASSP)}, pages 4087--4091. IEEE, 2014.

\bibitem{chen2015query}
G.~Chen, C.~Parada, and T.~N. Sainath.
\newblock Query-by-example keyword spotting using long short-term memory networks.
\newblock In {\em 2015 IEEE international conference on acoustics, speech and signal processing (ICASSP)}, pages 5236--5240. IEEE, 2015.

\bibitem{chu2021call}
H.~Chu, Y.~Wang, R.~Ju, Y.~Jia, H.~Wang, M.~Li, and Q.~Deng.
\newblock Call for help detection in emergent situations using keyword spotting and paralinguistic analysis.
\newblock In {\em Companion Publication of the 2021 International Conference on Multimodal Interaction}, pages 104--111, 2021.

\bibitem{ganin2016domain}
Y.~Ganin, E.~Ustinova, H.~Ajakan, P.~Germain, H.~Larochelle, F.~Laviolette, M.~March, and V.~Lempitsky.
\newblock Domain-adversarial training of neural networks.
\newblock {\em Journal of machine learning research}, 17(59):1--35, 2016.

\bibitem{gong2021ast}
Y.~Gong, Y.-A. Chung, and J.~Glass.
\newblock {AST: Audio Spectrogram Transformer}.
\newblock In {\em Proc. Interspeech 2021}, pages 571--575, 2021.

\bibitem{hoy2018alexa}
M.~B. Hoy.
\newblock Alexa, siri, cortana, and more: an introduction to voice assistants.
\newblock {\em Medical reference services quarterly}, 37(1):81--88, 2018.

\bibitem{hsu2021hubert}
W.-N. Hsu, B.~Bolte, Y.-H.~H. Tsai, K.~Lakhotia, R.~Salakhutdinov, and A.~Mohamed.
\newblock Hubert: Self-supervised speech representation learning by masked prediction of hidden units.
\newblock {\em IEEE/ACM Transactions on Audio, Speech, and Language Processing}, 29:3451--3460, 2021.

\bibitem{huang2021query}
J.~Huang, W.~Gharbieh, H.~S. Shim, and E.~Kim.
\newblock Query-by-example keyword spotting system using multi-head attention and soft-triple loss.
\newblock In {\em ICASSP 2021-2021 IEEE International Conference on Acoustics, Speech and Signal Processing (ICASSP)}, pages 6858--6862. IEEE, 2021.

\bibitem{hubballi2014false}
N.~Hubballi and V.~Suryanarayanan.
\newblock False alarm minimization techniques in signature-based intrusion detection systems: A survey.
\newblock {\em Computer Communications}, 49:1--17, 2014.

\bibitem{jeong2022cochlscene}
I.-Y. Jeong and J.~Park.
\newblock Cochlscene: Acquisition of acoustic scene data using crowdsourcing.
\newblock In {\em 2022 Asia-Pacific Signal and Information Processing Association Annual Summit and Conference (APSIPA ASC)}, pages 17--21. IEEE, 2022.

\bibitem{kim2019query}
B.~Kim, M.~Lee, J.~Lee, Y.~Kim, and K.~Hwang.
\newblock Query-by-example on-device keyword spotting.
\newblock In {\em 2019 IEEE automatic speech recognition and understanding workshop (ASRU)}, pages 532--538. IEEE, 2019.

\bibitem{kim2024stethoscope}
J.-W. Kim, S.~Bae, W.-Y. Cho, B.~Lee, and H.-Y. Jung.
\newblock Stethoscope-guided supervised contrastive learning for cross-domain adaptation on respiratory sound classification.
\newblock In {\em ICASSP 2024-2024 IEEE International Conference on Acoustics, Speech and Signal Processing (ICASSP)}, pages 1431--1435. IEEE, 2024.

\bibitem{kim2024repaugment}
J.-W. Kim, M.~Toikkanen, S.~Bae, M.~Kim, and H.-Y. Jung.
\newblock Repaugment: Input-agnostic representation-level augmentation for respiratory sound classification.
\newblock {\em arXiv preprint arXiv:2405.02996}, 2024.

\bibitem{kirandevraj2022generalized}
R.~Kirandevraj, V.~K. Kurmi, V.~P. Namboodiri, and C.~Jawahar.
\newblock Generalized keyword spotting using asr embeddings.
\newblock In {\em 23rd Annual Conference of the International Speech Communication Association, INTERSPEECH 2022}, pages 126--130. ISCA, 2022.

\bibitem{kong2020panns}
Q.~Kong, Y.~Cao, T.~Iqbal, Y.~Wang, W.~Wang, and M.~D. Plumbley.
\newblock Panns: Large-scale pretrained audio neural networks for audio pattern recognition.
\newblock {\em IEEE/ACM Transactions on Audio, Speech, and Language Processing}, 28:2880--2894, 2020.

\bibitem{liu2021rnn}
Z.~Liu, T.~Li, and P.~Zhang.
\newblock Rnn-t based open-vocabulary keyword spotting in mandarin with multi-level detection.
\newblock In {\em ICASSP 2021-2021 IEEE International Conference on Acoustics, Speech and Signal Processing (ICASSP)}, pages 5649--5653. IEEE, 2021.

\bibitem{mazumder2021few}
M.~Mazumder, C.~Banbury, J.~Meyer, P.~Warden, and V.~J. Reddi.
\newblock Few-shot keyword spotting in any language.
\newblock {\em arXiv preprint arXiv:2104.01454}, 2021.

\bibitem{mo2021encoder}
T.~Mo and B.~Liu.
\newblock Encoder-decoder neural architecture optimization for keyword spotting.
\newblock {\em arXiv preprint arXiv:2106.02738}, 2021.

\bibitem{ng2022i2cr}
D.~Ng, J.~Q. Yip, T.~Surana, Z.~Yang, C.~Zhang, Y.~Ma, C.~Ni, E.~S. Chng, and B.~Ma.
\newblock I2cr: Improving noise robustness on keyword spotting using inter-intra contrastive regularization.
\newblock In {\em 2022 Asia-Pacific Signal and Information Processing Association Annual Summit and Conference (APSIPA ASC)}, pages 605--611. IEEE, 2022.

\bibitem{nishu2023matching}
K.~Nishu, M.~Cho, and D.~Naik.
\newblock Matching latent encoding for audio-text based keyword spotting.
\newblock {\em arXiv preprint arXiv:2306.05245}, 2023.

\bibitem{park2019specaugment}
D.~S. Park, W.~Chan, Y.~Zhang, C.-C. Chiu, B.~Zoph, E.~D. Cubuk, and Q.~V. Le.
\newblock Specaugment: A simple data augmentation method for automatic speech recognition.
\newblock {\em Interspeech 2019}, 2019.

\bibitem{radford2023robust}
A.~Radford, J.~W. Kim, T.~Xu, G.~Brockman, C.~McLeavey, and I.~Sutskever.
\newblock Robust speech recognition via large-scale weak supervision.
\newblock In {\em International conference on machine learning}, pages 28492--28518. PMLR, 2023.

\bibitem{reddy2019scalable}
C.~K. Reddy, E.~Beyrami, J.~Pool, R.~Cutler, S.~Srinivasan, and J.~Gehrke.
\newblock A scalable noisy speech dataset and online subjective test framework.
\newblock 2019.

\bibitem{sacchi2019open}
N.~Sacchi, A.~Nanchen, M.~Jaggi, and M.~Cernak.
\newblock Open-vocabulary keyword spotting with audio and text embeddings.
\newblock In {\em INTERSPEECH 2019-IEEE International Conference on Acoustics, Speech, and Signal Processing}, 2019.

\bibitem{settle2017query}
S.~Settle, K.~Levin, H.~Kamper, and K.~Livescu.
\newblock Query-by-example search with discriminative neural acoustic word embeddings.
\newblock {\em arXiv preprint arXiv:1706.03818}, 2017.

\bibitem{shin2022learning}
H.-K. Shin, H.~Han, D.~Kim, S.-W. Chung, and H.-G. Kang.
\newblock Learning audio-text agreement for open-vocabulary keyword spotting.
\newblock {\em arXiv preprint arXiv:2206.15400}, 2022.

\bibitem{tan2019efficientnet}
M.~Tan and Q.~Le.
\newblock Efficientnet: Rethinking model scaling for convolutional neural networks.
\newblock In {\em International conference on machine learning}, pages 6105--6114. PMLR, 2019.

\bibitem{warden2018speech}
P.~Warden.
\newblock Speech commands: A dataset for limited-vocabulary speech recognition.
\newblock {\em arXiv preprint arXiv:1804.03209}, 2018.

\bibitem{yang2021superb}
S.-w. Yang, P.-H. Chi, Y.-S. Chuang, C.-I.~J. Lai, K.~Lakhotia, Y.~Y. Lin, A.~T. Liu, J.~Shi, X.~Chang, G.-T. Lin, et~al.
\newblock Superb: Speech processing universal performance benchmark.
\newblock {\em arXiv preprint arXiv:2105.01051}, 2021.

\bibitem{zhang2023u2}
A.~Zhang, P.~Zhou, K.~Huang, Y.~Zou, M.~Liu, and L.~Xie.
\newblock U2-kws: Unified two-pass open-vocabulary keyword spotting with keyword bias.
\newblock In {\em 2023 IEEE Automatic Speech Recognition and Understanding Workshop (ASRU)}, pages 1--8. IEEE, 2023.

\bibitem{zhang2017hello}
Y.~Zhang, N.~Suda, L.~Lai, and V.~Chandra.
\newblock Hello edge: Keyword spotting on microcontrollers.
\newblock {\em arXiv preprint arXiv:1711.07128}, 2017.

\end{thebibliography}

\end{document}